\documentstyle[aps,preprint,eqsecnum,floats,epsf]{revtex}
\tighten
\draft
\begin{document}

\preprint{MAD-NT-97-05}
\title{Green's Function for Nonlocal Potentials}
\author{A.~B. Balantekin\thanks{Electronic address:
        {\tt baha@nucth.physics.wisc.edu}},
        J.~F. Beacom\thanks{Electronic address:
        {\tt beacom@nucth.physics.wisc.edu}}, and
        M.~A. C\^andido Ribeiro\thanks{Electronic address:
        {\tt aribeiro@nucth.physics.wisc.edu}}}
\address{Department of Physics, University of Wisconsin\\
         Madison, Wisconsin 53706 USA}
\date{\today}
\maketitle

\begin{abstract}
The single-particle nuclear potential is intrinsically nonlocal.  In
this paper, we consider nonlocalities which arise from the many-body
and fermionic nature of the nucleus.  We investigate the effects of
nonlocality in the nuclear potential by developing the Green's
function for nonlocal potentials.  The formal Green's function
integral is solved analytically in two different limits of the
wavelength as compared to the scale of nonlocality.  Both results are
studied in a quasi-free limit.  The results illuminate some of the
basic effects of nonlocality in the nuclear medium.
\end{abstract}



\newpage
\section{General Aspects of Nonlocal Potentials}

A nonlocal term can be introduced to the non-relativistic
Schr\"odinger equation as follows:
\begin{equation}
-\frac{\hbar^2}{2m} \nabla^2 \Psi({\bf x}) +
\left[V_L({\bf x}) - E \right]\Psi({\bf x}) = 
- \int d^3 {\bf x'} \, V_{NL}({\bf x},{\bf x'}) \, \Psi({\bf x'})\,.
\label{nl1}
\end{equation}
Generally, such a nonlocal potential arises when one attempts to write
the equation of motion for a single particle interacting with a
many-body system.  One source of nonlocality is the coupling of the
elastic channel to other degrees of freedom.  The potential in the
elastic channel is modified by the contribution from excitation out of
the elastic channel at a point ${\bf x}$, propagation in an
intermediate state, and then de-excitation back to the elastic channel
at a point ${\bf x'}$.  This type of nonlocality is called Feshbach
nonlocality.  In addition, for a fermionic system, one must also
account for exchange effects, which require antisymmetrization of the
wave function between the projectile and the appropriate components of
the target.  This type of nonlocality is called Pauli nonlocality.

The form of $V({\bf x},{\bf x'})$ will not be considered in full
mathematical generality.  Nevertheless, this does not seem to be a
significant restriction for the physical theory.  We consider the
nonlocality to be characterized by a finite range $b$.  In particular,
we consider nonlocal potentials which limit to local potentials as the
range $b \rightarrow 0$.  This form allows sufficient richness in the
theory, but allows us to stay in contact with the familiar case of
local quantum mechanics.  This requirement excludes, for example,
nonlocal potentials of the form $V_1({\bf x}) V_2({\bf x'})$, which
have no classical analog as a limiting case in the sense
above\cite{Brink}.

In the development below, we consider a nonlocal potential operator
defined as
\begin{equation}
\hat V_{NL} = \int \int d^3{\bf x}\, d^3{\bf x'}\, 
V_{NL}({\bf x},{\bf x'})\, |{\bf x} \rangle \langle {\bf x'}|\,.
\label{PNL}
\end{equation}
For calculating the Green's function, we need the Fourier transform of
$V_{NL}({\bf x},{\bf x'})$, i.e.,
\begin{equation}
\tilde V_{NL}({\bf p},{\bf p'}) =
\langle {\bf p}| \hat V_{NL} | {\bf p'}\rangle\,.
\label{FT}
\end{equation}
Here and below a tilde is used to indicate the Fourier transform.
Calculation of these matrix elements reveals some other general
properties of the nonlocal potential.  To begin with, consider the
following form:
\begin{equation}
V_{NL}({\bf x},{\bf x'}) = V_1({\bf x},{\bf x'}) V_2({\bf x} - {\bf x'})\,.
\end{equation}
We require that the second term limit to a delta function as the range
of nonlocality $b \rightarrow 0$.  That insures that the potential has
a definite local limit.  The first term is allowed for now to have an
arbitrary dependence on ${\bf x}$ and ${\bf x'}$.  Then
\begin{eqnarray}
\tilde V_{NL}({\bf p},{\bf p'}) & = &
\frac{1}{(2 \pi \hbar)^3} \int \int d^3{\bf x}\, d^3{\bf x'}\, \nonumber\\
& \times &
\exp\left(\frac{i{\bf p'}\cdot {\bf x'}}{\hbar}\right)
\exp\left(-\frac{i{\bf p}\cdot {\bf x}}{\hbar}\right)
 V_1({\bf x},{\bf x'}) V_2({\bf x} - {\bf x'})\,.
\end{eqnarray}
We change variables to 
\begin{equation}
{\bf Y} = \frac{{\bf x} + {\bf x}'}{2}
\end{equation}
and
\begin{equation}
{\bf y} = {\bf x} - {\bf x}'
\end{equation}
so that the integrals can be partially separated:
\begin{eqnarray}
\tilde V_{NL}({\bf p},{\bf p'}) & = &
\frac{1}{(2 \pi \hbar)^3} \int d^3{\bf y}\, V_2({\bf y})
\exp\left(-\frac{i({\bf p}+{\bf p'})\cdot{\bf y}}{2\hbar}\right)\nonumber\\
& \times &
\int d^3{\bf Y}\, V_1({\bf Y},{\bf y})
\exp\left(-\frac{i({\bf p} - {\bf p'})\cdot {\bf Y}}{\hbar}\right)\,.
\label{FTsep}
\end{eqnarray}
In principle, the second integrand depends on ${\bf y}$.  Since we
assume that the potential is almost local, $V_2({\bf y})$ is very
sharply peaked in ${\bf y}$.  Then the ${\bf y}$-dependence of
$V_1({\bf Y},{\bf y})$ will not be seen at leading order in $b$.  In
that case, $V_1({\bf Y},{\bf y}) \approx V_1({\bf Y})$ alone.  This is
true to the extent that the range $b$ of the nonlocality is small on
the scale in which $V_1$ is changing.  For applications in nuclear
physics, where the width $b$ is of order 1 fm or less, this should be
a very reasonable approximation.  We have shown that the assumption of
a small nonlocality range $b$ motivates the following factorized form
\begin{equation}
V_{NL}({\bf x},{\bf x'}) = 
V_a\left(\frac{{\bf x} + {\bf x'}}{2}\right)
V_w\left({\bf x} - {\bf x'}\right)\,,
\label{fac}
\end{equation}
where $V_a$ (called $V_1$ above) is the ``average'' local value of the
nonlocal potential and $V_w$ (called $V_2$ above) governs the
``width'' of the nonlocality.  For a sufficiently small range of
nonlocality, the particular functional form of the width term should
not be too important.  What is important is that some nonlocality is
allowed, with a characteristic scale $b$.

If this factorized form is assumed, then the integrals in
Eq.~(\ref{FTsep}) separate, and then one finds for the Fourier
transform
\begin{equation}
\tilde V_{NL}({\bf p},{\bf p'}) =
U_a\left(\frac{{\bf p} + {\bf p'}}{2}\right)
U_w\left({\bf p} - {\bf p'}\right)\,,
\end{equation}
where, in the ${\bf p}$-representation, $U_a$ is the average potential
term, and $U_w$ is the width term.
These are given by:
\begin{equation}
U_a\left(\frac{{\bf p} + {\bf p'}}{2}\right) =
\tilde V_w\left({\bf x} - {\bf x'}\right)
\end{equation}
and
\begin{equation}
U_w\left({\bf p} - {\bf p'}\right) = 
\tilde V_a\left(\frac{{\bf x} + {\bf x'}}{2}\right)\,.
\end{equation}
This illuminates an interesting property of the factorized form.
Under the Fourier transform, the ``average'' term of the potential in
the ${\bf x}$-representation becomes the ``width'' term of the
potential in the ${\bf p}$-representation, and vice versa.  In
particular, there are the following correspondences (neglecting
factors):
\begin{eqnarray}
{\bf x}+{\bf x}' & \longleftrightarrow & {\bf p}-{\bf p}'\\
{\bf x}-{\bf x}' & \longleftrightarrow & {\bf p}+{\bf p}'\,.
\end{eqnarray}
These relations depend only on the particular factorization of
$V_{NL}$, and not upon the details of $V_a$ and $V_w$.

Let us consider a few special cases.  Consider first the width term
$V_w$.  For a small range of nonlocality $b$, $V_w$ is a narrow peak
(say Gaussian) in the ${\bf x}$-representation.  Then in the ${\bf
p}$-representation, the average term $U_a$ is a broad Gaussian.  Thus
nonlocality in the ${\bf x}$-representation leads to a modified
dispersion relation in the ${\bf p}$-representation.  The narrower the
nonlocality, the flatter the dispersion relation.  As $b \rightarrow
0$, the free-particle dispersion relation is recovered.  Now consider
the average term $V_a$.  If it is constant, then its Fourier transform
is proportional to a delta function, leading to a local potential in
the ${\bf p}$-representation.  To the extent that it is non-constant,
it leads to nonlocality in the ${\bf p}$-representation (via $U_w =
\tilde V_a$).

Typically, a Gaussian form is taken for $V_w$, with $V_w = V_w(|{\bf
x} - {\bf x'}|)$.  Other forms could be chosen, but this seems
adequate to describe the nuclear scattering data well \cite{PB}.
Besides having an evident local limit, this particular factorized form
meets some other requirements: $V_{NL}$ should be symmetric under the
exchange of ${\bf x}$ and ${\bf x'}$; and when $V_a = const.$, as in
infinite nuclear matter, $V_{NL}$ should be translation invariant.  In
this paper, we use the Gaussian form for $V_w$.  In Fig.~1, a
schematic matrix representation of the above discussion is given.


\section{Nonlocality in Nuclear Reactions}

As we have asserted in the previous section, nonlocality is a general
characteristic of a single-particle description of a many-body system.
It can be present, for example, in descriptions of systems in atomic
physics\cite{Mot}, condensed-matter physics\cite{JMZ}, and quantum
optics\cite{BM}.  Here, we are particularly interested in the specific
nonlocal potentials which can appear in nuclear reactions.  Below, we
consider the nonlocality due to channel coupling (Feshbach
nonlocality) and that due the fermionic character of the system (Pauli
nonlocality).

It is well known that a given nonlocal potential $V_{NL}({\bf x},{\bf
x'})$ can have a corresponding energy-dependent local equivalent
potential $V_{LE}({\bf x}, E)$\cite{SF}.  Such a local equivalent
potential gives the same cross section as the nonlocal potential.  If
the local equivalent potential also gives the same wave function, then
it is said to be trivially equivalent.

It was shown in a recent article\cite{Mar} that most of the observed
energy dependence in the inner region of the phenomenological local
equivalent potential arises from the Pauli nonlocality.  The Feshbach
nonlocality would then be related to the energy dependence at the
surface of that potential.  Exchange effects are important at
relatively short distances (inner region of the nuclear potential),
where the density is high.  Channel coupling effects are important at
relatively large distances (nuclear surface), where the coupling form
factors are peaked.  As an example of this kind of long range effect
we have, for example, the nuclear structure effects in Coulomb
excitation.  The relative contributions of the Feshbach and Pauli
nonlocalities can also be studied using the technique of
Ref.~\cite{Mack}.

We cannot give a rigorous proof that the factorized form for
$V_{NL}({\bf x},{\bf x'})$ in Eq.~(\ref{fac}) is completely general.
However, we will motivate this form for both the Feshbach and Pauli
nonlocalities below.


\subsection{Feshbach Nonlocality}

Recent theoretical and experimental developments in the study of 
fusion reactions below the Coulomb barrier led to a renewed interest in 
the coupled-channels formalism \cite{rmp}. 
For a long time it has been known that coupling to nonelastic channels
leads to a nonlocality in the elastic channel\cite{Fe}.  The
factorized nonlocal potential of Eq.~(\ref{fac}) is sometimes adopted
in channel coupling calculations, and seems to provide a good
description of the energy dependence of the nuclear potential.  The
generation of a nonlocality in the elastic channel due to channel
couplings has been described in detail in Ref.~\cite{Raw1}.  Here we
consider simplified example, using only two channels.  The coupled
Schr\"odinger equations in this case are given by:
\begin{equation}
\left[
\left(
\begin{array}{cc}
 \frac{\hat {\bf p}^2}{2 m} + V_L({\bf x}) & 0 \\
0 &  \frac{\hat {\bf p}^2}{2 m} + V_L({\bf x}) 
\end{array}
\right)
 + 
\left(
\begin{array}{cc}
0 & F({\bf x}) \\
F({\bf x}) & \varepsilon 
\end{array}
\right) 
\right]
\left(
\begin{array}{r}
\Psi_0({\bf x})\\
\Psi_1({\bf x}) 
\end{array}
\right)
=
E
\left(
\begin{array}{r}
\Psi_0({\bf x})\\
\Psi_1({\bf x}) 
\end{array}
\right)\,,
\label{coup}
\end{equation}
where $\Psi_0({\bf x})$ and $\Psi_1({\bf x})$ are the wave functions
of the ground and first excited states, respectively.  $V_L({\bf x})$
is a local potential, $F({\bf x})$ is the coupling potential and
$\varepsilon$ is the excitation energy to the first state.  These
differential equations can be solved by the Green's function method.
We can decouple these equations by eliminating either channel.  We
first solve for $\Psi_1({\bf x})$:
\begin{equation}
\Psi_1({\bf x}) =
\frac{2 m}{\hbar^2}
\int d^3{\bf x'}\;G({\bf x},{\bf x'})\;F({\bf x'})\;\Psi_0({\bf x'})\,,
\end{equation}
with the Green's function $G({\bf x},{\bf x'})$ is given by
\begin{equation}
G({\bf x},{\bf x'}) = \frac{\hbar^2}{2m}
\left \langle {\bf x}\left| \frac{1}{E - \hat{\bf p}^2/{2 m}-
V_L({\bf x}) - \varepsilon \pm i\eta}\right| {\bf x'} \right \rangle\,.
\end{equation}
Then the equation for $\Psi_0({\bf x})$ can be written as
\begin{equation}
\left[\frac{\hat {\bf p}^2}{2 m} + V_L({\bf x}) - E \right]\Psi_0({\bf
x}) = - \int d^3{\bf x'}\;F({\bf x})\; G({\bf x},{\bf x'})\;F({\bf
x'})\;\Psi_0({\bf x'})\,.
\label{nl2}
\end{equation}
Now that the $\Psi_1$ channel has been eliminated, this is an equation
for $\Psi_0$ alone.  By comparing Eq.~(\ref{nl2}) to the general form
in Eq.~(\ref{nl1}), we find that the channel coupling generates a
nonlocal potential
\begin{equation}
V_{NL}({\bf x},{\bf x'})= 
F({\bf x})\; G({\bf x},{\bf x'})\;F({\bf x'})\,.
\label{pnl}
\end{equation}
The nonlocality in $G({\bf x},{\bf x'})$ arises because the kinetic
energy is not diagonal in the ${\bf x},{\bf x'}$ representation.  This
and the elimination of all but the elastic channel leads to the
nonlocality in the elastic channel.

As a simple example, we can assume that $F({\bf x})$ has an
approximately exponential shape and that $V_L({\bf x})$ has a slow
variation.  Then we can write Eq.~(\ref{pnl}) as (keeping in mind that
in this situation we have $G({\bf x},{\bf x'}) \sim G({\bf x}-{\bf
x'})$\cite{FL}):
\begin{equation}
V_{NL}({\bf x},{\bf x'}) \approx
F({\bf x}+{\bf x'})\; G({\bf x} - {\bf x'})\,,
\end{equation}
which has the same form as Eq.~(\ref{fac}).


\subsection{Pauli Nonlocality}

The simple factorized form as given by Eq.~(\ref{fac}) has a clear
local limit when the nonlocality range $b$ goes to zero.  There are
good reasons to consider such a simple form for the nonlocal potential
$V_{NL}({\bf x},{\bf x'})$.  For example, Bauhoff {\it et
al.}\cite{Ba} examined Pauli nonlocality in nucleon-nucleus
interactions by using a folding approach.  Adopting a nucleon-nucleon
interaction written as the sum of a direct term and an exchange term,
these authors obtained a nonlocal nucleon-nucleus potential from the
microscopic description.  Aoki and Horiuchi\cite{AH}, using the
Resonating Group Method, verified that the microscopic interaction can
be decomposed as a sum of direct and exchange parts.  A comparison of
the results for the trivially and local equivalent potentials and
those potentials from the Perey-Buck analyses\cite{PB} was made in
Ref.~\cite{Raw2}.  It was shown that a large portion of the
nonlocality present in the Perey-Buck potential is due to exchange
effects.  The microscopic approach of Ref.~\cite{Ba} was found to
produce very similar results to those using the Perey-Buck factorized
form given above in Eq.~(\ref{fac}).

We conclude that the effects of the Pauli nonlocality in
nucleon-nucleus interactions can be successfully described by a
nonlocal potential given by Eq.~(\ref{fac}). The same conclusion can
be reached in the case of nucleus-nucleus interactions\cite{Mar,JJ}.


\subsection{Nonlocality and the Nuclear Dispersion Relation}

Here we note a consequence of allowing a nonlocal potential, as in
Eq.~(\ref{nl1}) to describe the nuclear interaction\cite{Fr,Fa}.  For
example, it is well known that the interaction described by
Eq.~(\ref{nl1}) is dependent on the energy of the interacting system.
In order to highlight the effects of nonlocality, we consider as an
example an infinite nuclear medium (average term $V_a$ constant) and a
Gaussian form for $V_w({\bf x} - {\bf x'})$.  Further, we neglect any
isolated local potential $V_L({\bf x})$.  We can then rewrite
Eq.~(\ref{fac}) as
\begin{equation} 
V_{NL}({\bf x},{\bf x'})= \frac{V_0}{\pi^{3/2}b^3} 
\exp\left[ -\frac{({\bf x} - {\bf x'})^2}{b^2}\right]\,.
\end{equation}
Here $b$ is the parameter describing the range of the nonlocal effects
and $V_0$ is the depth of the nuclear potential. In this way, we may
rewrite Eq. (\ref{nl1}) as
\begin{equation}
\left[ \frac{\hbar^2}{2m} \nabla^2 +
E \right]\Psi({\bf x}) = 
\frac{V_0}{\pi^{3/2}b^3} 
\int d^3 {\bf x'} \,\exp\left[ -\frac{({\bf x} - {\bf
x'})^2}{b^2}\right] \, \Psi({\bf x'})\,.
\end{equation}
We first change the integration variable to ${\bf s}$, defined by
$b{\bf s} = {\bf x'} - {\bf x}$.  Then using the relation $\exp(b{\bf
s}\cdot\nabla)\,\Psi({\bf x}) = \Psi ({\bf x}+b{\bf s})$, we can write
\begin{equation}
\left[ \frac{\hbar^2}{2m} \nabla^2 +
E \right]\Psi({\bf x}) = 
V_0 \exp\left(\frac{b^2}{4}\nabla^2\right) \,
\Psi({\bf x})\,.
\label{nl3}
\end{equation}
We replace $\Psi({\bf x})$ with its definition via the inverse Fourier
transform:
\begin{equation}
\Psi({\bf x}) = \left(\frac{1}{2\pi \hbar}\right)^{3/2} \int d^3{\bf p}
\,\exp\left(\frac{i{\bf p}\cdot {\bf x}}{\hbar}\right)\Phi({\bf p})\,.
\end{equation}
From this we obtain the momentum-space representation of Eq.~(\ref{nl3}):
\begin{equation}
\left(-\frac{{\bf p}^2}{2m} + E\right)\Phi({\bf p}) = 
V_0\exp\left(-\frac{{\bf p}^2 b^2}{4\hbar^2} \right)
\Phi({\bf p})\,.
\end{equation}
Defining ${\bf p} = \hbar {\bf k}$ and $k = |{\bf k}|$, the dispersion
relation for a nucleon moving in infinite nuclear matter is given by
\begin{equation}
\frac{\hbar^2 k^2}{2m} =
E - V_0 \exp\left(-\frac{k^2 b^2}{4}\right)\,.
\label{nldisp}
\end{equation}
Equation~(\ref{nldisp}) shows the dependence of the nuclear potential
on momentum.  Note that the strength of the potential decreases
quickly with increasing momentum.  For a local interaction ($b = 0$),
we recover the standard dispersion relation:
\begin{equation}
\frac{\hbar^2 k^2}{2m} = E - V_0\,.
\label{ldisp}
\end{equation}

Since Eq.~(\ref{nldisp}) is a transcendental equation, it cannot
solved for general momenta.  For small momenta, the exponential can be
expanded in a Taylor series.  Below, only the two lowest orders of
this expansion are kept.  Thus we require that $k b \ll 1$, which
means the nucleon wavelength is much larger than the range of the
nonlocality.  That is reasonable for the small range of nonlocality
that we assume.  Then the solution to Eq. (\ref{nldisp}) can be
rewritten
\begin{equation}
\frac{\hbar^2 \kappa^2}{2m^*} = (E - V_0)\,,
\label{kappa}
\end{equation}
where the effective mass $m^*$ is given by
\begin{equation}
\frac{m^{\ast}}{m} = \frac{1}{1 - (m b^2 V_0 /2 \hbar^2)}\,.
\label{mstar}
\end{equation}
Note that we have changed $k$ to $\kappa$ above to emphasize the fact
that $m^{\ast}$ is used in the definition of the latter.  Since $V_0$
is negative, $m^* < m$.  The nonlocal interaction between the nucleon
and the nuclear matter changes the dispersion relation, lowering the
effective mass.  In terms of the dispersion relation, a nucleon with a
mass $m$ and an nonlocal interaction is approximately equivalent to a
nucleon with a mass $m^*$ and a local interaction.


\section{Development of the Nonlocal Green's Function}

\subsection{General Development}

In this section, we study the Green's function for a nonlocal
potential and present our main results.  This allows a unifying
approach to various limiting cases, as will be shown below.  Studies
of the general properties of nonlocal potentials and some limits
thereof have been made in other formalisms\cite{Peierls,bul1}.  These
analytic studies are important in part because of the difficult nature
of solving these systems numerically (but note a new approach in
Ref.~\cite{Raw3}).  The Green's function may also be useful in
calculating the transmission coefficient in heavy-ion collisions,
where a nonlocal potential modifies the usual result\cite{Tunnel}.

We begin the development by defining the Green's operator for a
nonlocal potential as
\begin{equation}
\hat G_{NL} = \frac{1}{E - \hat H_{NL} \pm i\eta}\,,
\end{equation}
where
\begin{equation}
\hat H_{NL} = \frac{\hat {\bf p}^2}{2 m} + \hat V_{NL}\,,
\end{equation}
and $\hat V_{NL}$ is the nonlocal potential operator given by
Eq.~(\ref{PNL}).  An additional local potential will be considered in
a later section.

The Green's function in the ${\bf x}$-representation is given by
\begin{equation}
G_{NL}(E;{\bf x}_f,{\bf x}_i) \equiv
\frac{\hbar^2}{2 m}\langle{\bf x}_f\mid \hat G_{NL}\mid 
{\bf x}_i\rangle = \frac{\hbar^2}{2 m}\left\langle{\bf x}_f\left| 
\frac{1}{E-\hat H_{NL} \pm i\eta}\right| {\bf x}_i\right\rangle\,.
\label{sub}
\end{equation}
This can be rewritten as
\begin{eqnarray}
G_{NL}(E;{\bf x}_f,{\bf x}_i)&=&\frac{\hbar^2}{2 m} \int \int
d^3{\bf p}\;d^3{\bf p}' \nonumber \\ &\times& \langle{\bf x}_f\mid {\bf
p}\rangle \left\langle{\bf p}\left| \frac{1}{E-\hat
H_{NL}\pm i\eta}\right| {\bf p}'\right\rangle\langle{\bf p}'\mid 
{\bf x}_i\rangle
\label{ket}
\end{eqnarray}  
with
\begin{equation}
\langle{\bf x}_f\mid {\bf p}\rangle = 
\frac{\exp(i{\bf p}\cdot {\bf x}_f/\hbar)}{(2 \pi \hbar)^{3/2}}\,.
\end{equation}
To proceed further with Eq.~(\ref{ket}), we adopt a factorized form
for the nonlocal potential $V_{NL}$ as in Eq.~(\ref{fac}).

In order to understand the effects of nonlocality on the Green's
function, we consider a simple nonlocal potential.  We take the local
or average term in this potential to be constant.  For now, the form
of the nonlocal or width term will be left unspecified.  Thus
\begin{equation}
V_{NL}({\bf x},{\bf x'}) = 
V_a\left(\frac{{\bf x} + {\bf x'}}{2}\right)
V_w\left({\bf x} - {\bf x'}\right) =
V_0\, V_w\left({\bf x} - {\bf x'}\right)\,,
\end{equation}
where $V_0$ is a negative constant, and $V_w$ is peaked at ${\bf x} =
{\bf x'}$.  As discussed above, taking $V_a$ constant forces $U_w =
\tilde{V}_a$ to be a delta function, so that the potential is local in
the ${\bf p}$-representation.  That allows one of the integrals in
each of Eq.~(\ref{FT}) and Eq.~(\ref{ket}) to be collapsed.  Then
Eq.~(\ref{FT}) can be written as
\begin{equation}
\langle{\bf p}\mid \hat V_{NL}\mid{\bf p}'\rangle = 
\delta^3({\bf p}-{\bf p}')\, V_0\,
\int d^3{\bf y}\, V_w({\bf y}) 
\exp\left[-i\;{\bf y}\cdot {\bf p}/2 \hbar\right]\,,
\end{equation}
where ${\bf y} = {\bf x} - {\bf x'}$.  Then
\begin{equation}
\left\langle{\bf p}\left|\frac{1}{E-\hat H_{NL}\pm i\eta}\right|
{\bf p}'\right\rangle =
\frac{\delta^3({\bf p}-{\bf p}')}
{E - {\bf p}^2/2 m - V_0\, \tilde{V}_w({\bf p})\pm i \eta}\,,
\end{equation}
where a tilde indicates the Fourier transform.  Using these results,
Eq.~(\ref{ket}) can be written
\begin{equation}
G_{NL}(E;{\bf x}_f,{\bf x}_i)= \frac{\hbar^2}{2m}\int\;d^3{\bf
p}\;\frac{1}{(2 \pi\hbar)^3}\frac{\exp[{i{\bf p}\cdot ({\bf x}_f-{\bf
x}_i)/\hbar}]}{E-{\bf p}^2/2 m - V_0\,\tilde{V}_w({\bf p})\pm i \eta}
\label{green}\,. 
\end{equation}

For definiteness, we now choose for the nonlocal or width term
\begin{equation}
V_w({\bf y}) = \frac{1}{b^3 \pi^{3/2}}\;
\exp\left(-\frac{{\bf y}^2}{b^2}\right)\,.
\label{Vy}
\end{equation}
As noted above, this form has a simple local limit when $b \rightarrow
0$.  This implies for the Fourier transform
\begin{equation}
\tilde{V}_w({\bf p}) =
\exp\left(-\frac{{\bf p}^2 b^2}{4\hbar^2}\right)\,,
\label{Vp}
\end{equation} 
and thus for Eq.~(\ref{green})
\begin{equation}
G_{NL}(E;{\bf x}_f,{\bf x}_i)= \frac{\hbar^2}{2 m}\int d^3{\bf
p}\;\frac{1}{(2 \pi\hbar)^3}\;\frac{\exp{[i{\bf p}\cdot ({\bf x}_f-{\bf
x}_i)/\hbar}]}{E- {\bf p}^2 /2 m -V_0 \exp({-{\bf p}^2 b^2/4\hbar^2})
\pm i \eta}\,.
\label{green1}
\end{equation}
Defining new variables via $E = \hbar^2 k^2/2 m$ and ${\bf p} =
\hbar{\bf q}$ (and $q = |{\bf q}|$), this becomes
\begin{eqnarray}
G_{NL}(E;{\bf x}_f,{\bf x}_i) & = & 
\frac{1}{(2 \pi)^2}\int_0^{\infty}dq\, q^2\int_{-1}^{+1}d(\cos \theta)
\nonumber \\ &\times& \frac{\exp\left({i q |{\bf x}_f-{\bf x}_i|\cos
\theta}\right)}{k^2-q^2- \left({2 m}V_0 /\hbar^2\right)\exp({-q^2 b^2/4})
\pm i \eta}\,.
\end{eqnarray}
Making the integration over $\cos \theta$, we obtain
\begin{eqnarray}
G_{NL}(E;{\bf x}_f,{\bf x}_i) &=&
\frac{1}{8 \pi^2}\frac{i}{|{\bf x}_f-{\bf x}_i|}
\int_{-\infty}^{\infty}dq \,q \nonumber \\
&\times&
\frac{\left(\exp\left({i q |{\bf x}_f-{\bf x}_i|}\right)
-\exp\left({-i q |{\bf x}_f-{\bf x}_i|}\right)\right)}
{q^2+\alpha \exp({-\beta q^2})-(k^2\pm i \eta)}\,,
\label{integ}
\end{eqnarray}
where
\begin{equation}
\alpha = \frac{2 m V_0}{\hbar^2}
\label{alpha}
\end{equation}
and
\begin{equation}
\beta = \frac{b^2}{4}\,.
\label{beta}
\end{equation}
Equation~(\ref{integ}) is the starting point for the limits of the
subsequent sections.  To make the integration over $q$ in
Eq.~(\ref{integ}), we use the method of residues.  This requires that
we identify the poles of the integrand in the complex-$q$ plane.
Unfortunately, the equation
\begin{equation}
q^2 + \alpha \exp\left({-\beta q^2}\right) - (k^2\pm i \eta)=0
\label{pole}
\end{equation}
is a transcendental equation, and the roots cannot be found in the
general case.  Below, we consider two limiting approximations
involving $\beta$ in which the roots of this equation can be found and
the integral in Eq.~(\ref{integ}) done analytically.  In Fig.~2, the
problem of finding these roots is indicated schematically.


\subsection{Digression on the Poles of the Integrand}

The integral in Eq.~(\ref{integ}) is completely specified by the
residues at the roots in $q$ of Eq.~(\ref{pole}).  We show here how to
develop expansions for these roots.  In two cases of interest, these
expansions can be truncated with minimal (and calculable) error.  The
errors on the integral results can be calculated using the known
errors on the roots.  Since the final results are based on controlled
approximations, it is easy to determine where to truncate the
expansions.

Defining $y = (q/k)^2$, $A = \alpha/k^2$, and $B = \beta k^2$,
Eq.~(\ref{pole}) can be written in a completely dimensionless form as
\begin{equation}
y + A e^{-B y} - 1 = 0\,.
\end{equation}
For convenience, the $i\eta$ term has been dropped.  From Fig.~2, one
can see that there are exactly two roots on the real axis.  We ignore
any possible complex roots as they would not lead to pure propagating
wave solutions in the Green's function.  We exclude the case $|A| \gg
1$; otherwise the value of $A$ is unrestricted.

\underline{Limit of $B = \beta k^2 \ll 1$}:  Rewrite Eq.~(\ref{pole}) as
\begin{eqnarray}
y & = & 1 - A e^{-B y} \nonumber \\
y & = & (1 - A)
+ A\left(B y - \frac{1}{2} B^2 y^2 + \ldots \right) \nonumber \\
(1 - A B)\, y
& = & (1 - A) + A \left(-\frac{1}{2} B^2 y^2 + \ldots \right)\,.
\end{eqnarray}
These are exact if all terms are kept.  For $B \ll 1$, the latter term
is small.  The corrections can be solved for by perturbation,
assuming the form
\begin{equation}
(1 - A B)\, y = y_0 + B y_1 + B^2 y_2 \ldots\,,
\end{equation}
and solving order by order.  Then $y_0 = 1 - A$, and the first
correction (the $y_1$ term) vanishes.  Then Eq.~(\ref{pole}) can be
written as
\begin{equation}
(1 - \alpha\beta)\, q^2 - (k^2 - \alpha \pm i\eta)
+ {\cal O}\left(k^2 (\beta k^2)^2\right) = 0\,.
\label{Bl1limit}
\end{equation}
This is a direct replacement for the denominator of Eq.~(\ref{integ})
near the poles; note that we did not divide out the factor $(1 -
\alpha\beta)$.

\underline{Limit of $B = \beta k^2 \gg 1$}:  Rewrite Eq.~(\ref{pole}) as
\begin{eqnarray}
y & = & 1 - A e^{-B y} \nonumber \\
y & = & \left(1 - A e^{-B}\right) - A \left(e^{-B y} - e^{-B}\right)\,.
\end{eqnarray}
These are exact.  For $B \gg 1$, $y \simeq 1$, so the latter term is
small.  The corrections can be solved for by iteration to yield a power
series in $e^{-B}$.  Note that $B \gg 1$ and $e^{-B} \ll 1$, so that
$B e^{-B} \ll 1$.  Then Eq.~(\ref{pole}) can be written as
\begin{equation}
q^2 - \left(k^2 - \alpha \exp(-\beta k^2) \pm i\eta\right) + 
{\cal O}\left(k^2 \exp(-2 \beta k^2)\right) = 0\,.
\label{Bg1limit}
\end{equation}
This is a direct replacement for the denominator of Eq.~(\ref{integ})
near the poles.


\subsection{Low-Momentum (or Effective Mass) Limit}

If $\beta k^2 \ll 1$, then the denominator of of Eq.~(\ref{integ}) can
be rewritten as in Eq.~(\ref{Bl1limit}) and so Eq.~(\ref{integ}) becomes
\begin{eqnarray}
G_{NL}(E;{\bf x}_f,{\bf x}_i) & = & 
\frac{1}{8 \pi^2}\frac{i}{|{\bf x}_f-{\bf x}_i|} 
\frac{m^{\ast}}{m} \int_{-\infty}^{\infty} dq\,q \nonumber \\ 
& \times &
\frac{\left[\exp\left({i q |{\bf x}_f-{\bf x}_i|}\right)
-\exp\left({-i q |{\bf x}_f-{\bf x}_i|}\right)\right]}
{q^2 - (\kappa^2 \pm i \eta)}\,, 
\label{inte}
\end{eqnarray}
where
\begin{equation}
\frac{m^{\ast}}{m}=\frac{1}{1 - \alpha \beta}\,,
\end{equation}
is the effective mass and 
\begin{equation}
\kappa^2 = \frac{k^2-\alpha}{1 - \alpha \beta}\,,
\label{kap2}
\end{equation}
is a new wave number that is equal to the wave number of the nucleon
in a constant local potential with the effective mass $m^*$.  By
construction, these expressions agree with those found in
Eqs.~(\ref{kappa}) and (\ref{mstar}).  Note that the corrections to
Eq.~(\ref{Bl1limit}) have been dropped.

The integral in Eq.~(\ref{inte}) is done by the method of residues,
with the contour closed in the upper-half or lower-half complex plane
as necessary.  We obtain the following expression for $G_{NL}(E;{\bf
x}_f,{\bf x}_i)$ in the limit $\beta k^2 \ll 1$:
\begin{equation}
G_{NL}(E;{\bf x}_f,{\bf x}_i) = 
-\frac{1}{4 \pi}
\frac{\exp\left({\pm i\kappa
| {\bf x}_f-{\bf x}_i|}\right)}{| {\bf x}_f-{\bf x}_i|}
\left(\frac{m^{\ast}}{m} \right)\,.
\label{g1} 
\end{equation}
Nonlocality modifies the mass (making the term in parentheses
different from unity) and modifies the wave number (which is also
modified by the constant average term in the potential.)  The local
limit is immediately recovered as $m^{\ast} \rightarrow m$ and $\kappa
\rightarrow k$.  On the other hand, if additional orders in
Eq.~(\ref{Bl1limit}) are retained, then the $m^*/m$ prefactor is
unchanged and only $\kappa$ is modified.

Note that this Green's function is an eigenstate of $p^2/\hbar^2 =
-\nabla^2$ with eigenvalue $\kappa^2$.  Using the
definition of $\kappa^2$ in Eq.~(\ref{kap2}),
$k^2 = 2mE/\hbar^2$ and the definitions in
Eqs.~(\ref{alpha}) and (\ref{beta}), this can be written as
\begin{equation}
\frac{p^2}{2 m^*} = \frac{\hbar^2 \kappa^2}{2 m^*} = E - V_0\,.
\end{equation}
This check of the integration by direct differentiation returns the
correct dispersion relation at the order of approximation used in
this case.


\subsection{High-Momentum Limit}

If $\beta k^2 \gg 1$, then the denominator of of Eq.~(\ref{integ}) can
be rewritten as in Eq.~(\ref{Bg1limit}) and so Eq.~(\ref{integ}) becomes
\begin{eqnarray}
G_{NL}(E;{\bf x}_f,{\bf x}_i) & = & 
\frac{1}{8 \pi^2}\frac{i}{|{\bf x}_f-{\bf x}_i|} 
\int_{-\infty}^{\infty} dq\,q \nonumber \\ 
& \times &
\frac{\left[\exp\left({i q |{\bf x}_f-{\bf x}_i|}\right)
-\exp\left({-i q |{\bf x}_f-{\bf x}_i|}\right)\right]}
{q^2 - (k^2 - \alpha \exp(-\beta k^2) \pm i \eta)}\,.
\end{eqnarray}
This is again done by residues, and is found to be
\begin{eqnarray}
G_{NL}(E;{\bf x}_f,{\bf x}_i) =
-\frac{1}{4 \pi}
\frac{\exp\left({\pm i\sqrt{k^2 - \alpha\exp\left(-\beta k^2\right)}\,
|{\bf x}_f-{\bf x}_i|}\right)}{| {\bf x}_f-{\bf x}_i|}\,.
\label{g2}
\end{eqnarray}
The small term $\exp(-\beta k^2)$ damps the effect of the constant
average potential.  Other than the modifed wavenumber, the result in
Eq.~(\ref{g2}) is exactly the usual free-particle Green's function.
Note that $k$ is the free-particle wave number, and is not modified by
the constant average potential.  If additional orders in
Eq.~(\ref{Bg1limit}) are retained, then the form is unchanged but the
wavenumber is corrected by powers of $\exp(-\beta k^2)$.  

Note that this Green's function is an eigenstate of $p^2/\hbar^2 =
-\nabla^2$ with eigenvalue $k^2 - \alpha\exp(-\beta k^2)$.  Using the
definition $k^2 = 2mE/\hbar^2$ and the definitions in
Eqs.~(\ref{alpha}) and (\ref{beta}), this can be written as
\begin{equation}
\frac{p^2}{2 m} = E - V_0 \exp(-m b^2 E/2\hbar^2)\,.
\end{equation}
This check of the integration by direct differentiation returns the
correct dispersion relation at the order of approximation used in
this case.


\subsection{Quasi-Free Limit}

Above, we simplified the integral by considering the limits of $B =
\beta k^2 \ll 1$ and $B = \beta k^2 \gg 1$, respectively.  The value
of $A = \alpha/k^2$ was not specified, except for excluding the case
$|A| \gg 1$.  The case $|A| \ll 1$ was allowed, and to consider its
consequences it is sufficient to apply this limit to the previous
results.  It is not necessary to develop a new expansion for the roots
or to do another integral.

In the low-momentum case, we took $B \ll 1$ and the result was
Eq.~(\ref{g1}).  If $|A| \ll 1$, then $|A| B = |\alpha| \beta \ll 1$,
so that $m^*/m \rightarrow 1$ and $\kappa^2 \rightarrow k^2 - \alpha$.
\begin{equation}
G_{NL}(E;{\bf x}_f,{\bf x}_i) \rightarrow
-\frac{1}{4 \pi}
\frac{\exp\left({\pm i\sqrt{k^2 - \alpha}\,
|{\bf x}_f-{\bf x}_i|}\right)}{| {\bf x}_f-{\bf x}_i|}\,.
\label{g3}
\end{equation}
This is of the same form as the free-particle Green's function, with
the wave number modified by the constant average potential.  This is a
local limit, since the scale $\beta$ has disappeared.

In the high-momentum case, we took $B \gg 1$ and the result was
Eq.~(\ref{g2}).  In the case $|A| \ll 1$, the result is unchanged
at the lowest nontrivial order:
\begin{eqnarray}
G_{NL}(E;{\bf x}_f,{\bf x}_i) \rightarrow
-\frac{1}{4 \pi}
\frac{\exp\left({\pm i\sqrt{k^2 - \alpha\exp\left(-\beta k^2\right)}\,
|{\bf x}_f-{\bf x}_i|}\right)}{| {\bf x}_f-{\bf x}_i|}\,.
\label{g4}
\end{eqnarray}
This is also of the same form as the free-particle Green's function,
again with the wave number modified by a term depending on the
interaction.  In this case, the nonlocality still appears as part of
the interaction.  Note that it would appear that these two Green's
functions share a common limit if in the second we take $\beta k^2 \ll
1$.  Strictly speaking, this should not be allowed since the
derivation assumed $\beta k^2 \gg 1$.


\subsection{Comparison to Data}

From Eqs.~(\ref{g1}), (\ref{g2}), (\ref{g3}), and (\ref{g4}) we can
see some of the basic effects of nonlocality in the nuclear medium.
Nonlocality, in the low-momentum limit, changes the constant mass of
the system to an effective mass, which depends on the range of
nonlocality and the strength of the potential.  This effect, in the
limit $|\alpha| \beta \ll 1$, is negligible. In the high-momentum
limit, nonlocality acts on the potential in a more complex way.
However, this limit is satisfied in many cases of interest to nuclear
physics\cite{Ch}. In this situation, the static strength of the
potential changes, under the effects of the nonlocality, to a
exponentially energy-dependent potential, e.g., the equivalent local
potential. Chamon {\it et al.}\cite{Ch} verified such a dependence of
the equivalent local potential on the energy for elastic scattering of
many different systems at high energies.

We can define an energy-dependent local equivalent potential from
Eq.~(\ref{g2}) as
\begin{eqnarray}
V(E)=V_0 \exp(-\beta k^2) \nonumber
\end{eqnarray}
or, substituting the expressions for $\beta$ and $k^2$, as 
\begin{equation}
V(E)=V_0 \exp\left(-\frac{m b^2}{2 \hbar^2}E\right)\,.
\label{ept}
\end{equation}
As a test of our results, we compare Eq.~(\ref{ept}) to the values of
the inner region of the phenomenological potential extracted from
data\cite{Mar} for the systems $^{12}$C + $^{12}$C and $^{12}$C +
$^{16}$O. In these comparisons, we adopted for the range of the
nonlocality $b = 0.14$ fm and $b = 0.11$ fm for the first and second
systems, respectively.  These values of $b$ are extracted from the
data, and are in good agreement with those from a single-folding
model.  The values of $V_0$ also come from the data.  For more details
about the data used, see Ref.~\cite{Mar} and references therein.

In the upper panel of Fig.~3, $|V(E)|$ versus $E$ for the system
$^{12}$C + $^{12}$C is shown. The circles represent the values of the
phenomenological potential at $r = 4$ fm extracted from the data and
the solid line represents the result of Eq.~(\ref{ept}). In the lower
panel of Fig.~3 the same quantities for the system $^{12}$C + $^{16}$O
is shown.  For the $^{12}$C + $^{12}$C system, $|\alpha| \beta =
0.085$; for the $^{12}$C + $^{16}$O system, $|\alpha| \beta = 0.075$.
In both cases there is an impressive concordance between the extracted
and predicted values for the potential.


\subsection{Born Series to Include a Local Potential}

Referring back to Eq.~(\ref{ket}), we note that in order to proceed we
needed to collapse the double integral over ${\bf p}$ and ${\bf p'}$.
To do that, we took the interaction to be local in the ${\bf
p}$-representation, i.e., assumed a constant average potential in the
${\bf x}$-representation.  (In the usual development of the Green's
function for a local interaction, any constant potential is defined
away by shifting the energy.)  One way to generalize the usual Green's
function integral is to allow a constant potential with a nonlocality,
as done here.  This introduces a $V({\bf p})$, and the remaining
single integral can be done in some cases.

To introduce an additional local potential, one can proceed as
follows.  We define a total Green's function by
\begin{equation}
\hat G = \frac{1}{E-\hat H_{NL}-\hat V_L \pm i\eta}
\end{equation}
where
\begin{equation}
\hat H_{NL} = \frac{\hat {\bf p}^2}{2 m} +\hat V_{NL}\,.
\end{equation}
Here, $\hat V_{NL}$ is a nonlocal potential operator, as given by
Eq.~(\ref{PNL}), and $\hat V_{L}$ is a local potential operator.
Then 
\begin{equation}
\hat G = \hat G_{NL} + \hat G_{NL}\;\hat V_L\; \hat G\,,
\label{ser}
\end{equation}
with
\begin{equation}
\hat G_{NL}=\frac{1}{E-\hat H_{NL} \pm i\;\eta}\,.
\end{equation}
The Born series can be developed in the usual way:
\begin{equation}
\hat G = \hat G_{NL} + \hat G_{NL}\;\hat V_L\;\hat G_{NL} +
\hat G_{NL}\;\hat V_L\;\hat G_{NL}\;\hat V_L\;\hat G_{NL} + . . . \,.
\end{equation}
This Born series for the operator $\hat G$ may or may not converge
depending on the value of the energy of the system or on the strength
of the local potential.  The order by order contributions to the
matrix element $\langle{\bf x}_f\mid \hat G\mid {\bf x}_i\rangle$ can
then be developed once the potential is specified.


\section{Conclusions}

In this article we presented a Green's function approach to the study
of nonlocal nuclear potentials.  We showed that Green's functions can
be used to investigate various limits in a unified way.  In the
low-momentum limit the effects of the nonlocal interaction can be
accounted for with an effective mass and a wavenumber modified by the
strength of the potential and range of the nonlocality.  In the
high-momentum limit the Green's function for the nonlocal potential
also differs from the local one by a modified wavenumber.  In this
case, it is a function of the free wave number, i.e., the nonlocal
potential is written in terms of an energy-dependent local equivalent
potential.  This behavior is consistent with the results recently
obtained by Chamon, {\it et al.} \cite{Ch}, who showed that as energy
goes up the equivalent local potential for the nucleus-nucleus
interactions become shallower.  It would also be interesting to
generalize the Green's function method to multidimensional systems
where the effective mass becomes a tensor \cite{bul2}.

In the two main limits which we considered, we kept only the leading
order results.  However, in both cases we have shown how to easily
generalize these results to any arbitrary order.

The results in this paper are readily applicable to nuclear matter. We
expect our formalism to be useful to incorporate Pauli non-locality
effects in the study of neutron stars.


\section*{ACKNOWLEDGMENTS}

We thank L.C. Chamon for illuminating discussions, and K. Hagino for
pointing out and discussing Ref.~\cite{Raw1} with us.  This work was
supported in part by the U.S. National Science Foundation Grant No.\
PHY-9605140 at the University of Wisconsin, and in part by the
University of Wisconsin Research Committee with funds granted by the
Wisconsin Alumni Research Foundation.  M.A.C.R.\ acknowledges the
support of Funda\c c\~ao de Amparo \`a Pesquisa do Estado de S\~ao
Paulo (Contract No.\ 96/3240-5).



\newpage
\centerline{\bf Figure Captions}

\bigskip
FIG. 1.  Schematic matrix representations of a nonlocal potential.
The solid dots indicate nonzero matrix elements, and the vertical bars
indicate the relative magnitude of those elements.  In the matrix in
the top left, only representative elements are indicated.  In the top
row, the potential has a constant local term multiplied by a Gaussian
nonlocal term with a range $b$.  On the left, the potential is
represented in the ${\bf x},{\bf x'}$ basis; on the right, in the the
${\bf p}, {\bf p'}$ basis.  The two bases are related by a Fourier
transform (FT).  In the bottom row, we consider the local limit ($b =
0$) of the top row.

\bigskip
FIG. 2.  The problem of finding the root of the dispersion relation
(as a function of momentum $q$) is indicated schematically in the
figure.  The figure is symmetric about $q=0$.  The parabolic curve
indicates the kinetic energy term $q^2/2m$.  The Gaussian curve
indicates the nonlocal potential term $V_0 \exp(-q^2 b^2/4)$, with
$V_0 < 0$.  The roots are at the points where the curves cross.  When
$b q \ll 1$, the Gaussian is extremely broad and the roots are near $q
= \pm \sqrt{2 m (E + |V_0|)}/\hbar$.  When $b q \gg 1$, the Gaussian
is very narrow and the roots are near $q = \pm \sqrt{2 m E}/\hbar$.

\bigskip
FIG. 3.  The energy dependence of the local equivalent potential
$|V(E)|$ at $r=4$ fm in the range $10\leq E/A \leq 200$ MeV/nucleon.
The circles are data, and the solid line is the result of
Eq.~(\ref{ept}).  (See the text for details.)  The upper panel is for
$^{12}$C + $^{12}$C, and the lower panel is for $^{12}$C + $^{16}$O.

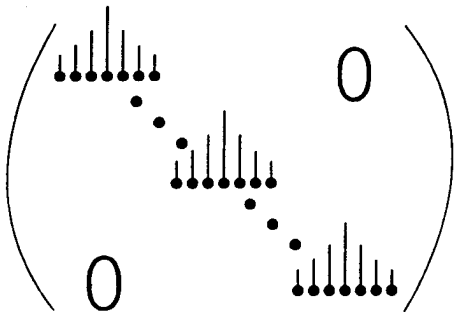
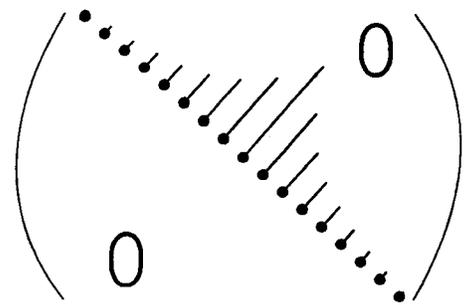
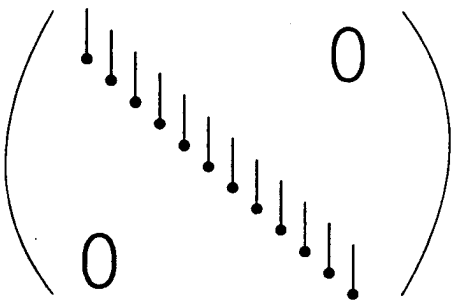
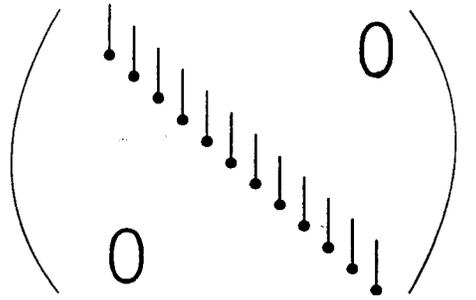
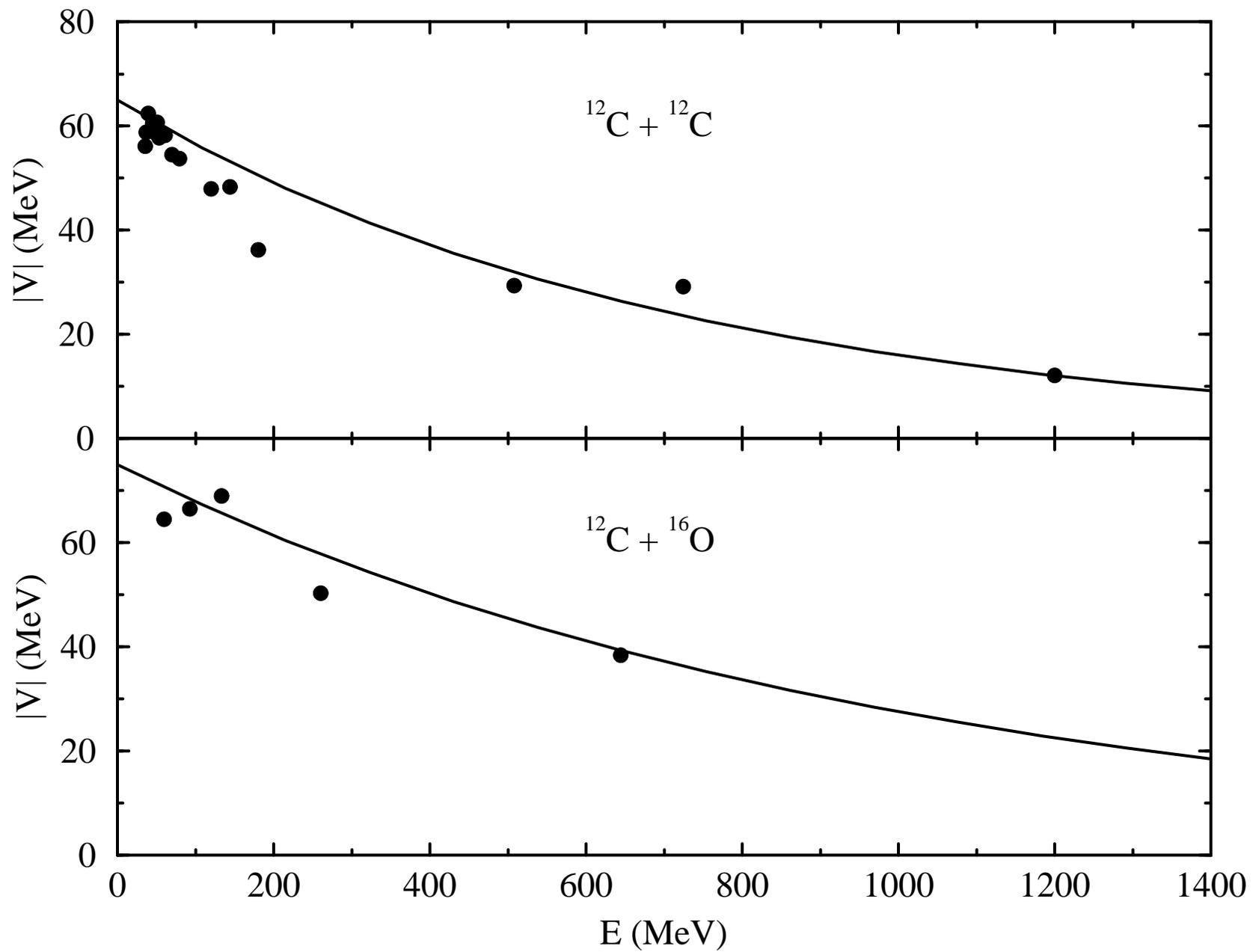
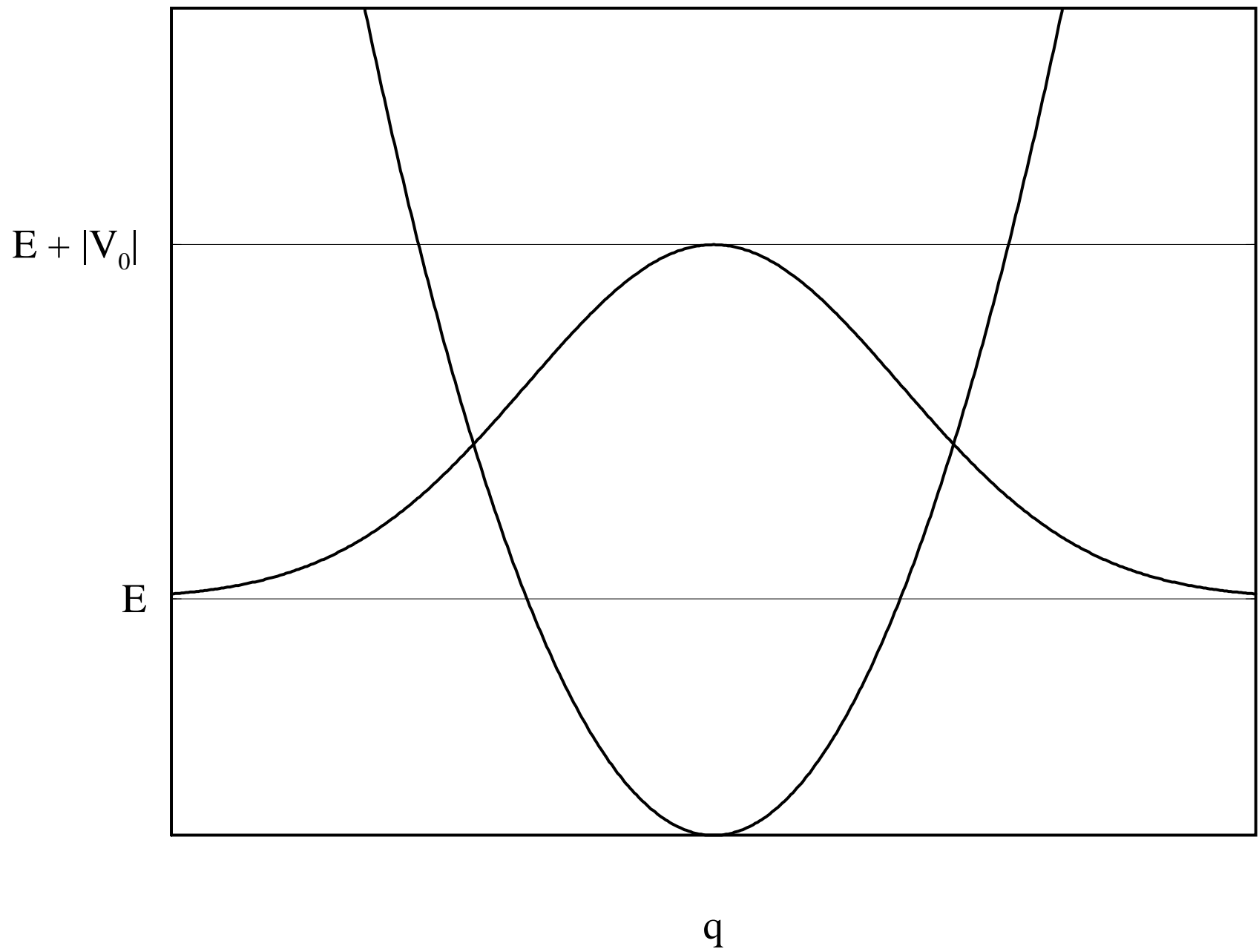
\end{document}